\documentstyle[sprocl]{article}

\catcode`\@=11\relax
\newwrite\@unused
\def\typeout#1{{\let\protect\string\immediate\write\@unused{#1}}}
\def\figurepath{[]}

\def\@nnil{\@nil}
\def\@empty{}
\def\@psdonoop#1\@@#2#3{}
\def\@psdo#1:=#2\do#3{\edef\@psdotmp{#2}\ifx\@psdotmp\@empty \else
    \expandafter\@psdoloop#2,\@nil,\@nil\@@#1{#3}\fi}
\def\@psdoloop#1,#2,#3\@@#4#5{\def#4{#1}\ifx #4\@nnil \else
       #5\def#4{#2}\ifx #4\@nnil \else#5\@ipsdoloop #3\@@#4{#5}\fi\fi}
\def\@ipsdoloop#1,#2\@@#3#4{\def#3{#1}\ifx #3\@nnil 
       \let\@nextwhile=\@psdonoop \else
      #4\relax\let\@nextwhile=\@ipsdoloop\fi\@nextwhile#2\@@#3{#4}}
\def\@tpsdo#1:=#2\do#3{\xdef\@psdotmp{#2}\ifx\@psdotmp\@empty \else
    \@tpsdoloop#2\@nil\@nil\@@#1{#3}\fi}
\def\@tpsdoloop#1#2\@@#3#4{\def#3{#1}\ifx #3\@nnil 
       \let\@nextwhile=\@psdonoop \else
      #4\relax\let\@nextwhile=\@tpsdoloop\fi\@nextwhile#2\@@#3{#4}}
\def\psdraft{
        \def\@psdraft{0}
}
\def\psfull{
        \def\@psdraft{100}
}
\psfull
\newif\if@prologfile
\newif\if@postlogfile
\newif\if@noisy
\def\pssilent{
        \@noisyfalse
}
\def\psnoisy{
        \@noisytrue
}
\psnoisy
\newif\if@bbllx
\newif\if@bblly
\newif\if@bburx
\newif\if@bbury
\newif\if@height
\newif\if@width
\newif\if@rheight
\newif\if@rwidth
\newif\if@clip
\newif\if@verbose
\def\@p@@sclip#1{\@cliptrue}

\def\@p@@sfile#1{\def\@p@sfile{null}%
                \openin1=#1
                \ifeof1\closein1%
                       \openin1=\figurepath#1
                        \ifeof1\typeout{Error, File #1 not found}
                        \else\closein1
                            \edef\@p@sfile{\figurepath#1}%
                        \fi%
                 \else\closein1%
                       \def\@p@sfile{#1}%
                 \fi}
\def\@p@@sfigure#1{\def\@p@sfile{null}%
                \openin1=#1
                \ifeof1\closein1%
                       \openin1=\figurepath#1
                        \ifeof1\typeout{Error, File #1 not found}
                        \else\closein1
                            \def\@p@sfile{\figurepath#1}%
                        \fi%
                 \else\closein1%
                       \def\@p@sfile{#1}%
                 \fi}

\def\@p@@sbbllx#1{
                \@bbllxtrue
                \dimen100=#1
                \edef\@p@sbbllx{\number\dimen100}
}
\def\@p@@sbblly#1{
                \@bbllytrue
                \dimen100=#1
                \edef\@p@sbblly{\number\dimen100}
}
\def\@p@@sbburx#1{
                \@bburxtrue
                \dimen100=#1
                \edef\@p@sbburx{\number\dimen100}
}
\def\@p@@sbbury#1{
                \@bburytrue
                \dimen100=#1
                \edef\@p@sbbury{\number\dimen100}
}
\def\@p@@sheight#1{
                \@heighttrue
                \dimen100=#1
                \edef\@p@sheight{\number\dimen100}
}
\def\@p@@swidth#1{
                \@widthtrue
                \dimen100=#1
                \edef\@p@swidth{\number\dimen100}
}
\def\@p@@srheight#1{
                \@rheighttrue
                \dimen100=#1
                \edef\@p@srheight{\number\dimen100}
}
\def\@p@@srwidth#1{
                \@rwidthtrue
                \dimen100=#1
                \edef\@p@srwidth{\number\dimen100}
}
\def\@p@@ssilent#1{ 
                \@verbosefalse
}
\def\@p@@sprolog#1{\@prologfiletrue\def\@prologfileval{#1}}
\def\@p@@spostlog#1{\@postlogfiletrue\def\@postlogfileval{#1}}
\def\@cs@name#1{\csname #1\endcsname}
\def\@setparms#1=#2,{\@cs@name{@p@@s#1}{#2}}
\def\ps@init@parms{
                \@bbllxfalse \@bbllyfalse
                \@bburxfalse \@bburyfalse
                \@heightfalse \@widthfalse
                \@rheightfalse \@rwidthfalse
                \def\@p@sbbllx{}\def\@p@sbblly{}
                \def\@p@sbburx{}\def\@p@sbbury{}
                \def\@p@sheight{}\def\@p@swidth{}
                \def\@p@srheight{}\def\@p@srwidth{}
                \def\@p@sfile{}
                \def\@p@scost{10}
                \def\@sc{}
                \@prologfilefalse
                \@postlogfilefalse
                \@clipfalse
                \if@noisy
                        \@verbosetrue
                \else
                        \@verbosefalse
                \fi
}
\def\parse@ps@parms#1{
                \@psdo\@psfiga:=#1\do
                   {\expandafter\@setparms\@psfiga,}}
\newif\ifno@bb
\newif\ifnot@eof
\newread\ps@stream
\def\bb@missing{
        \if@verbose{
                \typeout{psfig: searching \@p@sfile \space  for bounding box}
        }\fi
        \openin\ps@stream=\@p@sfile
        \no@bbtrue
        \not@eoftrue
        \catcode`\%=12
        \loop
                \read\ps@stream to \line@in
                \global\toks200=\expandafter{\line@in}
                \ifeof\ps@stream \not@eoffalse \fi
                \@bbtest{\toks200}
                \if@bbmatch\not@eoffalse\expandafter\bb@cull\the\toks200\fi
        \ifnot@eof \repeat
        \catcode`\%=14
}       
\catcode`\%=12
\newif\if@bbmatch
\def\@bbtest#1{\expandafter\@a@\the#1
\long\def\@a@#1
\long\def\bb@cull#1 #2 #3 #4 #5 {
        \dimen100=#2 bp\edef\@p@sbbllx{\number\dimen100}
        \dimen100=#3 bp\edef\@p@sbblly{\number\dimen100}
        \dimen100=#4 bp\edef\@p@sbburx{\number\dimen100}
        \dimen100=#5 bp\edef\@p@sbbury{\number\dimen100}
        \no@bbfalse
}
\catcode`\%=14
\def\compute@bb{
                \no@bbfalse
                \if@bbllx \else \no@bbtrue \fi
                \if@bblly \else \no@bbtrue \fi
                \if@bburx \else \no@bbtrue \fi
                \if@bbury \else \no@bbtrue \fi
                \ifno@bb \bb@missing \fi
                \ifno@bb \typeout{FATAL ERROR: no bb supplied or found}
                        \no-bb-error
                \fi
                \count203=\@p@sbburx
                \count204=\@p@sbbury
                \advance\count203 by -\@p@sbbllx
                \advance\count204 by -\@p@sbblly
                \edef\@bbw{\number\count203}
                \edef\@bbh{\number\count204}
}
\def\in@hundreds#1#2#3{\count240=#2 \count241=#3
                     \count100=\count240        
                     \divide\count100 by \count241
                     \count101=\count100
                     \multiply\count101 by \count241
                     \advance\count240 by -\count101
                     \multiply\count240 by 10
                     \count101=\count240        
                     \divide\count101 by \count241
                     \count102=\count101
                     \multiply\count102 by \count241
                     \advance\count240 by -\count102
                     \multiply\count240 by 10
                     \count102=\count240        
                     \divide\count102 by \count241
                     \count200=#1\count205=0
                     \count201=\count200
                        \multiply\count201 by \count100
                        \advance\count205 by \count201
                     \count201=\count200
                        \divide\count201 by 10
                        \multiply\count201 by \count101
                        \advance\count205 by \count201
                     \count201=\count200
                        \divide\count201 by 100
                        \multiply\count201 by \count102
                        \advance\count205 by \count201
                     \edef\@result{\number\count205}
}
\def\compute@wfromh{
                \in@hundreds{\@p@sheight}{\@bbw}{\@bbh}
                \edef\@p@swidth{\@result}
}
\def\compute@hfromw{
                \in@hundreds{\@p@swidth}{\@bbh}{\@bbw}
                \edef\@p@sheight{\@result}
}
\def\compute@handw{
                \if@height 
                        \if@width
                        \else
                                \compute@wfromh
                        \fi
                \else 
                        \if@width
                                \compute@hfromw
                        \else
                                \edef\@p@sheight{\@bbh}
                                \edef\@p@swidth{\@bbw}
                        \fi
                \fi
}
\def\compute@resv{
                \if@rheight \else \edef\@p@srheight{\@p@sheight} \fi
                \if@rwidth \else \edef\@p@srwidth{\@p@swidth} \fi
}
\def\compute@sizes{
        \compute@bb
        \compute@handw
        \compute@resv
}
\def\psfig#1{\vbox {
        \ps@init@parms
        \parse@ps@parms{#1}
        \compute@sizes
        \ifnum\@p@scost<\@psdraft{
                \if@verbose{
                        \typeout{psfig: including \@p@sfile \space }
                }\fi
                \special{ps::[begin]    \@p@swidth \space \@p@sheight \space
                                \@p@sbbllx \space \@p@sbblly \space
                                \@p@sbburx \space \@p@sbbury \space
                                startTexFig \space }
                \if@clip{
                        \if@verbose{
                                \typeout{(clip)}
                        }\fi
                        \special{ps:: doclip \space }
                }\fi
                \if@prologfile
                    \special{ps: plotfile \@prologfileval \space } \fi
                \special{ps: plotfile \@p@sfile \space }
                \if@postlogfile
                    \special{ps: plotfile \@postlogfileval \space } \fi
                \special{ps::[end] endTexFig \space }
                \vbox to \@p@srheight true sp{
                        \hbox to \@p@srwidth true sp{
                                \hss
                        }
                \vss
                }
        }\else{
                \vbox to \@p@srheight true sp{
                \vss
                        \hbox to \@p@srwidth true sp{
                                \hss
                                \if@verbose{
                                        \@p@sfile
                                }\fi
                                \hss
                        }
                \vss
                }
        }\fi
}}
\def\psglobal{\typeout{psfig: PSGLOBAL is OBSOLETE; use psprint -m instead}}
\catcode`\@=12\relax

\bibliographystyle{unsrt} 

\arraycolsep1.5pt

\def\Journal#1#2#3#4{{#1} {\bf #2}, #3 (#4)}

\def\NCA{\em Nuovo Cimento}
\def\NIM{\em Nucl. Instrum. Methods}
\def\NIMA{{\em Nucl. Instrum. Methods} A}
\def\NPB{{\em Nucl. Phys.} B}
\def\PLB{{\em Phys. Lett.}  B}
\def\PRL{\em Phys. Rev. Lett.}
\def\PRD{{\em Phys. Rev.} D}
\def\ZPC{{\em Z. Phys.} C}

\def\st{\scriptstyle}
\def\sst{\scriptscriptstyle}
\def\mco{\multicolumn}
\def\epp{\epsilon^{\prime}}
\def\vep{\varepsilon}
\def\ra{\rightarrow}
\def\ppg{\pi^+\pi^-\gamma}
\def\vp{{\bf p}}
\def\ko{K^0}
\def\kb{\bar{K^0}}
\def\al{\alpha}
\def\ab{\bar{\alpha}}
\def\be{\begin{equation}}
\def\ee{\end{equation}}
\def\bea{\begin{eqnarray}}
\def\eea{\end{eqnarray}}
\def\CPbar{\hbox{{\rm CP}\hskip-1.80em{/}}}

\def\nub{\overline{\nu}}
\def\num{\nu_{\mu}}
\def\numb{\overline{\nu}_{\mu}}
\def\nue{\nu_{e}}
\def\nueb{\overline{\nu}_{e}}

\begin{document}

\title{RECENT QCD RESULTS IN $\nu-N$ DEEP-INELASTIC-SCATTERING AT CCFR/NUTEV}

\author{JAEHOON YU\\
(for the CCFR/NuTeV Collaborations)}

\address{MS309, FNAL, P.O.Box 500, Batavia,\\ IL 60510, USA\\E-mail: yu@fnal.gov} 


\maketitle\abstracts{ 
We present recent QCD results in 
      $\nu-N$ scattering at the Fermilab CCFR/NuTeV experiments.
We present the latest Next-to-Next-Leading order strong coupling constant, 
      $\alpha_{s}$, extracted from Gross-Llewellyn-Smith sum rule.
The value of $\alpha_{s}$ from this measurement, at the
      mass of Z boson, is 
      $\alpha_{s}^{NNLO}(M_{Z}^{2})=0.114^{+0.009}_{-0.012}$.
We also present a preliminary result of the CCFR $F_{2}$ at 
      large-$x$.
This measurement of $F_{2}$ is compared to other experiments 
      and various models that provide nuclear effects.
We discuss the previous strange sea quark measurement from
      the CCFR experiment and the prospects for improvements of 
      the measurement in the NuTeV experiment.
}

\section{Introduction}
Neutrino-nucleon ($\nu$-N) deep inelastic scattering (DIS) experiments provide 
        a good testing field for Quantum Chromo-Dynamics (QCD), the theory of 
        strong interactions.
The $\nu$-N DIS experiments probe the structure of nucleons and provide an
        opportunity to test QCD evolutions and to extract the strong coupling
        constant, $\alpha_{s}$.
They are complementary measurements to charged lepton DIS experiments of
        nucleon structure functions.
The advantage of $\nu$-N DIS measurements over the charged lepton experiments
        is that $\nu$-N experiments can measure 
        both $F_{2}(x,Q^{2})$ and $xF_{3}(x,Q^2)$ due to pure V-A nature.
The $\nu-N$ differential cross sections are written, in terms of structure 
        functions:
\begin{eqnarray}\label{eq:nun-xsec}
\frac{d^{2}\sigma^{\nu(\overline{\nu})}}{dxdy}
=\frac{G_{F}^{2}ME_{\nu}}{\pi}
\left[\left(1-y-\frac{Mxy}{2E_{\nu}}
+\frac{y^{2}}{2}\frac{1+4M^{2}x^{2}/Q^{2}}{1+R(x,Q^{2})}\right)
F_{2}^{\nu(\overline{\nu})} \right.  \nonumber \\
\left.  \pm \left(y-\frac{y^{2}}{2}\right)
xF_{3}^{\nu(\overline{\nu})}\right]
\end{eqnarray}
where $R(x,Q^{2})=\sigma_{L}/\sigma_{T}$,
the ratio of longitudinal to transverse absorption cross sections.
The structure functions $F_{2}(x,Q^{2})$ and $xF_{3}(x,Q^{2})$ are extracted
        from fitting Eq.~\ref{eq:nun-xsec} to the 
        measured differential cross sections. 

In this paper, we present a next-to-next leading order (NNLO) 
        determination of 
        $\alpha_{s}$ from the Gross-Llewellyn-Smith (GLS) sum rule 
        and a large-$x$ structure function measurement from the CCFR 
        experiment.
We also discuss the measurements of strange sea quark distributions from 
        the CCFR
        experiment and the expected improvements of this measurement in
        NuTeV.

\section{The Experiments}\label{ss:detector}
CCFR (Columbia-Chicago-Fermilab-Rochester)/NuTeV experiment is a 
        $\nu-N$ DIS experiment at the Tevatron in Fermilab.
The CCFR experiment used a broad momentum beam of mixed neutrinos 
        ($\nu_{\mu}$) and antineutrinos ($\overline{\nu}_{\mu}$) from 
        decays of the secondary pions and kaons, resulting from
        interactions of 
        800GeV primary protons on a Beryllium-Oxide (BeO) target.
The NuTeV experiment, the successor of CCFR, used separate beams
        of $\nu_{\mu}$ or $\overline{\nu}_{\mu}$ during a given running 
        period, using a Sign-Selected-Quadrupole-Train (SSQT)~\cite{ex:ssqt}.

The CCFR/NuTeV detector~\cite{ex:det} consists of two major components : 
        target calorimeter and muon spectrometer.
The target calorimeter is an iron-liquid-scintillator sampling calorimeter,
        interspersed with drift chambers to provide track information of
        the muons resulting from charged-current (CC) interactions where
        a charged weak boson (${\rm W^{+}}$ or ${\rm W^{-}}$) is 
        exchanged between
        the $\nu_{\mu}$ ($\overline{\nu}_{\mu}$) and the parton.
The calorimeter provides dense material in the path of $\nu$ 
        ($\overline{\nu}$) to increase the rate of neutrino interactions.
The hadron energy resolution of the calorimeter is measured from the
        test beam and is found to be :
$\sigma/E_{Had}=(0.847\pm 0.015)/\sqrt{E_{Had}(GeV)}+(0.30\pm0.12)/E_{Had}$.

The muon spectrometer is located immediately downstream of the target 
        calorimeter and consists of three toroidal magnets and five drift 
        chamber stations to provide accurate measurements of muon momenta.
The momentum resolution of the spectrometer ($\sigma/p_{\mu}$) is 
        approximately 10.1\% and the angular resolution is 
        $\theta_{\mu}=0.3+60/p_{\mu} ({\rm mrad})$.

\section{$\alpha_{s}$ from Gross-Llewellyn-Smith Sum Rule}\label{ss:gls}

Once the structure function $xF_{3}$ is extracted, one can use 
        the Gross-Llewellyn-Smith (GLS) sum rule~\cite{th:gls_sum},
        which states that $\int xF_{3} (dx/x)$ is the total number of 
        valence quarks in a nucleon, up to QCD corrections, 
        to extract the strong coupling constant, $\alpha_{s}$.
Since in leading order (LO), the structure function $xF_{3}$ is 
        $xq-x\overline{q}$, the valence quark distributions, integrating 
        $xF_{3}$ over $x$ yields total number of valence quarks, 3.

Since the GLS sum rule is a fundamental prediction of QCD and the integral 
        only depends on valence quark distributions, $\alpha_{s}$ can be
        determined without being affected by less well known gluon 
        distributions.
Moreover, since there are sufficient number of measurements of $xF_{3}$ 
        in a wide range of 
        $Q^{2}$, one can measure $\alpha_{s}$ as a function of $Q^{2}$ in
        low $Q^{2}$ where the values of $\alpha_{s}$ depends most 
        on $\Lambda_{QCD}$.

With next-to-next-to-leading order QCD corrections~\cite{th:nnlo_gls}, 
        the GLS integral takes the form :
\begin{equation}\label{eq:gls_int}
\int_{0}^{1} xF_{3}(x,Q^{2})\frac{dx}{x} =
3 \left(1 - \frac{\alpha_{s}}{\pi} - a(n_{f})(\frac{\alpha_{s}}{\pi})^{2}
-b(n_{f})(\frac{\alpha_{s}}{\pi})^{3}\right) -\Delta HT
\end{equation}
where the term $\Delta HT$ is the corrections from higher-twist effects.
The higher-twist correction term, $\Delta HT$, is predicted to be significant
        in some models~\cite{th:gls_ht}, while others~\cite{th:ht_others}
        predict negligibly small corrections.
We take $\Delta HT$ as one half the largest model prediction with the 
        associated error covering the full range 
        ($\Delta HT=(0.15\pm0.15)/Q^{2}$).

\begin{figure}
\centerline{\psfig{figure=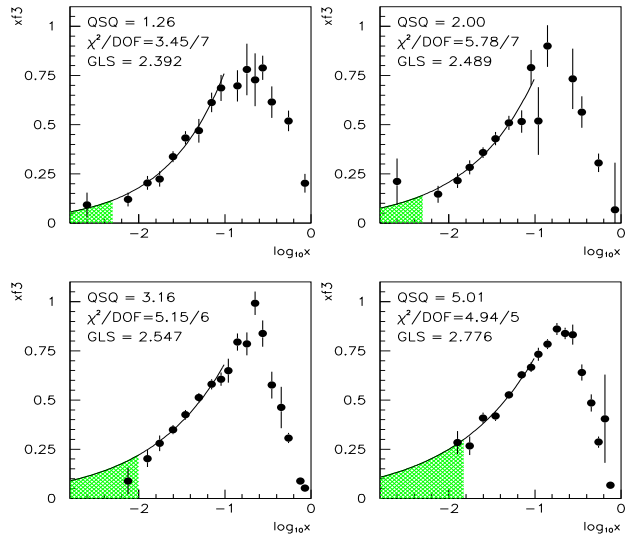,width=3.5in,height=3.0in}}
\caption[]{$xF_{3}$ vs $x$ for four lowest $Q^{2}$ bins.   The solid circles
represent the CCFR $xF_{3}$ data and the inverse triangle represent the
data from other experimental measurements.}
\label{fg:xf3_vs_x}
\end{figure}
In order to perform the integration of $xF_{3}$ in the entire ranges of $x$,
        we use data from other $\nu$-N DIS experiments 
        (WA59, WA25, SKAT, FNAL-E180, and BEBC-Gargamelle~\cite{ex:allxf3}),
        to cover large-$x$ ranges that are not covered by CCFR due to 
        geometric and kinematic acceptances.
The CCFR data has a minimum $x$ of roughly $x=0.002Q^{2}$.
To extrapolate below this kinematic limit, we fit $xF_{3}$ to a power 
        law ($Ax^{B}$) using all the data points in $x<0.1$.

Since the data points do not populate the region $x>0.5$ finely, 
        it is necessary to interpolate between the data points.
Thus, we use the shapes of the charged lepton DIS $F_{2}$ in $x>0.5$,
        because the shapes of $xF_{3}$ and $F_{2}$ should be the same in
        this region of $x$ due to negligible contribution from sea quarks.
The charged lepton $F_{2}$ data are corrected for nuclear 
        effects~\cite{ex:emc-effect} before obtaining the shapes.
Systematic uncertainties due to extrapolation and interpolation take 
        into account the possible variations of the models in the low-$x$
        region and the resonance peaks in charged-lepton $F_{2}$ shapes
        in large-$x$ regions.

Figure~\ref{fg:xf3_vs_x} shows $xF_{3}$ in the four lowest $Q^{2}$ bins as 
        a function of $x$.
The solid circles represent the experimental $xF_{3}$ data,
The solid lines represent the power law fit used for extrapolation of $xF_{3}$
        outside the kinematic limit of CCFR.
The shaded area in each plot shows the region of $x$ where the extrapolations
        are used for the integration.

The $xF_{3}$ integrals are estimated in six $Q^{2}$ bins.
The results in each $Q^{2}$ bin are fit to the NNLO pQCD function
        and higher-twist effect in Eq.~\ref{eq:gls_int}.
This procedure yields a best fit of $\Lambda_{\overline{MS}}^{5,NNLO}=165$MeV.
Evolving this result to the mass of the Z boson, $M_{Z}$, in NNLO, this corresponds
        to the value of $\alpha_{s}$:
\begin{equation}
\alpha_{s}^{NNLO}(M_{Z}^{2})=0.114^{+0.005}_{-0.006}(stat.)
^{+0.007}_{-0.009}(syst.)\pm 0.005(theory)
\end{equation}
where systematic uncertainty includes the error in the fit and
        theory error represents the uncertainty due to higher twist 
        effects and higher order QCD contributions.
This result is consistent with
        the value extracted from the CCFR structure function 
        measurement~\cite{ex:ccfr-bill}.

Evolving to the mean value of $Q^{2}$ ($=3{\rm GeV^{2}}$) for this 
        analysis results in
$\alpha_{s}^{NNLO}(3{\rm GeV^{2}})=0.278\pm0.035\pm 0.05^{+0.035}_{-0.03}$.
If the higher-twist effect is neglected, the value of 
        $\alpha_{s}^{NNLO}(M_{Z}^{2})$ becomes 0.118.

\section{Large-$x$ Structure Functions}
In a simple parton model, the contributions of a single stationary nucleons
       vanish as $x$, the fractional parton momentum, approaches to 1.
However, the interplays between nucleons in a nuclear environment would 
       add a momentum to initial nucleon that interacts with the 
       neutrino, causing $x$ to be greater than 1.
This nuclear effect has been observed and measured in EMC experiment and is
       called the EMC effect~\cite{ex:emc-effect}.

There have been several models to describe this nuclear effect :
1) Fermi-gas model which adds Fermi momentum~\cite{th:fgas} and 
   a quasi-deutron effect~\cite{th:fgas-deut},
2) few nucleon correlation model~\cite{th:fnc} and 
3) the multi-quark cluster model~\cite{th:bag}, 
    which add exponentially falling $F_{2}$ distributions.

\begin{figure}[tpb]
\vskip -1in
\centerline{\psfig{figure=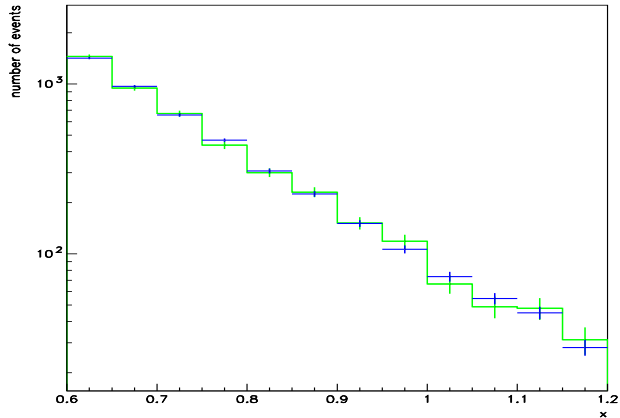,width=3.5in,height=2.5in}}
\vskip 0.3in
\caption[]{Number of events vs $x$ for all events in $Q^{2}>50{\rm GeV^{2}}$.
The histogram represent the data and the points represents the fit results.}
\label{fg:largex}
\end{figure}

In order to explore these various models, the CCFR experiment measured 
    large-$x$ cross sections in the kinematic regions $x>0.75$ and
    $Q^{2}>50 {\rm GeV^{2}}$.
This measured cross section is compared to a LO Monte Calro predictions
    that includes various detector smearing effects and incorporates
    various models separately.
Figure~\ref{fg:largex} shows the number of events as a function of the scaling 
     variable $x$ in the region $x>0.75$.
The structure function $F_{2}(x,Q^{2})$ is then fit to the functional 
    form~\cite{ex:bcdms-par}
    $F_{2}(x,Q^{2})=(1-x)^{a}(b+cx+dx^{2}+ex^{3})(Q^{2})^{f+gx}$
\footnote{a:2.5693, b=0.2739, c=3.0437, d=-5.5172, e=2.5790, f=-0.0303, 
and g=-0.2185} for $x<0.75$
    and $F_{2}(x,Q^{2})=F_{2}(x=0.75,Q^{2})e^{-s(x-0.75)}$ for $x>0.75$.
The resulting value of the parameter $s$ from 
     this fit is $s=8.3\pm0.7(stat)\pm0.3(syst)$.

We also investigated a possible $Q^{2}$ dependence of this parameter $s$ in
    two $Q^{2}$ bins, $50<Q^{2}<100~{\rm GeV^{2}}$ and 
    $100<Q^{2}<400~{\rm GeV^{2}}$.
The resulting values of the parameter $s$ in these two kinematic regions are
    $s=7.4\pm0.9$ for lower $Q^{2}$ bin and $s=8.7\pm0.9$ in the higher $Q^{2}$
    bin, showing only a weak $Q^{2}$ dependences.
Table~\ref{tb:largex} compares this result with other experimental measurements
    and models, demonstrating that the FNC prediction 
    and the 
\begin{table}[h]
\begin{center}
\caption[]{Comparisons of the exponent $s$ of CCFR with other predictions and
experimental measurements.}\label{tb:largex}
\begin{tabular}{|c|c|}\hline
Experiment/Predictions&$s$ \\ \hline\hline
CCFR & $8.3\pm 0.8$ \\ \hline
FNC Prediction & $8\sim 9$ \\ \hline
Baldin {\it et al.} Prediction & $\sim 6$ \\ \hline
SLAC E133 & $7\sim 8$ \\ \hline
BCDMS & $16.5\pm 0.5$ \\ \hline
\end{tabular}
\end{center}
\end{table}
SLAC E133 measurement of the parameter $s$ agree with
    our result, while the BCDMS result differs significantly.

From this analysis, we find that the models without nuclear effect do not 
      describe our data well.
Incorporating only Fermi gas and quasi-deutron model is 
      not enough to describe the excess of the data over $x>1$.
The FNC model describes the CCFR large-$x$ behavior well with the exponential
      fall off parameter of $s=8.3\pm0.9$.
This measurement indicates that the nucleon structure function $F_{2}$ at
      $x=1$ does not vanish but has a finite value~\cite{ex:vakili} of 
      $\sim 2\times 10^{-3}$.

\section{Strange Sea Measurements}
In $\nu-N$ DIS, the flavor changing weak CC interactions provide a clean 
      signature for a scattering off an $s$ quark ($s\rightarrow c$), 
      resulting in a pair of oppositely signed muons in the final state
      through the reaction 
 $\nu_{\mu}(\overline{\nu}_{\mu})+N\rightarrow \mu^{-}
(\mu^{+})+c(\overline{c})+X$ where 
 $c(\overline{c})$ subsqeuantly undergoes a semileptonic decay
   $c(\overline{c})\rightarrow s(\overline{s})+\mu^{+}(\mu^{-})
+\overline{\nu}_{\mu}(\nu_{\mu})$.
In contrast, detecting $s$ quark with NC interactions in a charged lepton
      DIS is convoluted with $s$ production and fragmentation of strange 
      mesons, requiring good particle identifications.
The nucleon structure function $F_{2}$ from $\nu-N$ DIS together with that
      measured from charged lepton DIS could also measure $s$ quark 
      distributions.
In addition, the difference of the $\nu-N$ structure function $xF_{3}$ between
      $\nu$ and $\overline{\nu}$ measures the nucleon strange quark contents
      as well.

CCFR experiment performed both LO~\cite{ex:ccfr_dimu-lo} 
       and NLO analyses~\cite{ex:ccfr_dimu}.
Since a neutrino from the semileptonic decay of the charm is involved 
       in the final state, the CCFR measured the
       distributions of visible physical quantities; $x_{vis}$, 
       $E_{vis}=E_{Had}+p_{\mu}^{1}+p+{\mu}^{2}$, 
       and $z_{vis}=\frac{p_{\mu}^{2}}{E_{Had}+p_{\mu}^{2}}$ 
       which is the fractional momentum taken by the
       charm meson fragmentation.
The measured distributions were compared to Monte Carlo predictions which
       incorporate detector acceptance and resolutions.

The Monte Carlo included
       the singlet ($xq_{SI}(x,\mu^{2})$), non-singlet 
      ($xq_{NS}(x,\mu^{2})=xq_{V}(x,\mu^2)$), and gluon parton
      distribution functions from CCFR structure function analysis assuming:
\begin{eqnarray}
xq_{V}(x,\mu^2)=xu_{V}(x,\mu^2)+xd_{V}(x,\mu^2)\\
xd_{V}(x,\mu^2)=A_{d}(1-x)xu_{V}(x,\mu^2)\\
x\overline{u}(x,\mu^2)=xu_{S}(x,\mu^2)\\
x\overline{d}(x,\mu^2)=xd_{S}(x,\mu^2)
\end{eqnarray}
The above assumes symmetric sea quark distributions.
We then parameterized the strange sea distribution to:
\begin{eqnarray}
xs(x,\mu^2)=A_{s}(1-x)^{\alpha}
\left[\frac{x\overline{u}(x,\mu^2)+x\overline{d}(x,\mu^2)}{2}\right]
\end{eqnarray}
where, $A_{s}$ is defined in terms of level and shape parameters,
 $\kappa$ and $\alpha$.
The level parameter $\kappa$ is defined as :
\begin{eqnarray}
\kappa=\frac{\int_{0}^{1}[xs(x,\mu^2)+x\overline{s}(x,\mu^2)]}
{\int_{0}^{1}[x\overline{u}(x,\mu^2)+x\overline{d}(x,\mu^2)]}
\end{eqnarray}

The LO prediction included a simple parton model which has tree level
       calculations for scattering of a $W$ off an $s$ quark, resulting in 
       a charm quark final state.
On the other hand, the NLO prediction included the ACOT 
         formalism~\cite{th:acot} whose 
         calculation is performed in a variable flavor ${\rm \overline{MS}}$ 
         scheme.
Specifically the NLO prediction included the Born and gluon-fusion diagrams.
In addition, since the charm quark is massive, the threshold effect was
         taken into account in the prediction via leading order slow 
         rescaling formalism where the scaling variable $x$ is replaced
         with $\xi$ defined as :
\begin{eqnarray}\label{eq:slow-xi}
\xi=x\left (1+\frac{m_{c}^{2}}{Q^{2}} \right)
\left( 1-\frac{x^{2}M^{2}}{Q^{2}} \right ),
\end{eqnarray}
where $m_{c}$ is the mass of the charm quark.

We then fit the Monte Carlo to data distributions of $x_{vis}$, 
          $E_{vis}$, and $z_{vis}$ for $\kappa$, $\alpha$, $B_{c}$ (the
          charmed meson semi-leptonic branching ratio), 
          $m_{c}$, and $\epsilon$ (fragmentation parameter),
          using the values of CKM martix elements $\mid V_{cd}\mid$ and 
          $\mid V_{cs}\mid$, taken from PDG~\cite{ex:pdg}.
\begin{table}[tpb]
\caption[]{Summary of the CCFR strange sea analysis.  The first set of the
errors in each item is statistical and the second set is the systematic 
uncertainties.}\label{ex:dimu}
\begin{center}
\begin{tabular}{|c|c|c|c|c|c|}\hline\hline
 &Fragmentation & $\kappa$ & $\alpha$ & $B_{c}$ & $m_{c}$(GeV) \\ \hline
 & Collins-Spiller & 0.477 & -0.02 & 0.1091 & 1.70 \\ 
NLO& $\epsilon = 0.81$ & $^{+0.046}_{-0.044}$& $^{+0.60}_{-0.54}$
& $^{+0.0082}_{-0.0074}$ & $\pm 0.17$\\
& $\pm 0.14$ &$^{+0.023}_{-0.024}$ & $^{+0.28}_{-0.26}$ 
& $^{+0.0063}_{-0.0051}$
& $^{+0.09}_{-0.08}$ \\ \hline
& Peterson & 0.468 & -0.05 & 0.1047 & 1.69 \\
NLO& $\epsilon = 0.20$ & $^{+0.061}_{-0.046}$ & $^{+0.46}_{-0.47}$
& $\pm 0.0076$ & $\pm 0.16$ \\
  &$\pm 0.04$ &$^{+0.024}_{-0.025} $ &$^{+0.28}_{-0.26}$
 & $^{+0.0065}_{-0.0052}$ &$^{+0.12}_{-0.10}$ \\ \hline
 & Peterson & 0.373 & 2.50 & 0.1050 & 1.31 \\ 
LO& $\epsilon = 0.20$ & $^{+0.048}_{-0.041}$
& $^{+0.60}_{-0.55}$
& $\pm 0.007$
& $^{+0.20}_{-0.22}$\\
 &$\pm 0.04$ &$\pm0.018 $ &$^{+0.36}_{-0.25}$ &$\pm 0.005$ 
 &$^{+0.12}_{-0.11}$ \\ \hline\hline
\end{tabular}
\end{center}
\end{table}

Table~\ref{ex:dimu} summarizes the results of this analysis, comparing
          various fragmentation functions at LO and NLO.
The result shows that the strange sea level parameter, $\kappa$, is in 
          qualitative agreement between LO and NLO analyses while the shape
          parameter, $\alpha$, is the same as other sea quarks in NLO but
          softer in LO.
The charm quark mass differs between NLO and LO, due presumably to 
          the fact that the parameter $m_{c}$ is not a pole mass but rather 
          a theoretical parameter that absorbs the lack of higher corrections.

Aside from the large statistical uncertainty, there were two major sources of 
         systematic uncertainties in performing 
         this measurement in CCFR.
First and foremost one is the identification of the secondary muons 
         resulting from the decay of charmed mesons.
Since the neutrino beam in CCFR was a mixture of $\num$ and $\numb$, the 
         experiment relied on transverse momentum of the muons relative to
         the neutrino beam direction, $P_{T}^{\mu}$, to distinguish
         the prompt muons from secondary muons.
The second systematic uncertainty is the LO slow rescaling formalism to take into account
         the charm threshold effect in theoretical predictions.

The experimental systematic uncertainty due to the identification of
         the secondary muon almost completely disappears
         in the NuTeV experiment due to the use of SSQT whose wrong sign
         neutrino contamination at a given mode is less than $10^{-3}$.
In addition, the NuTeV experiment is working on improving low energy muon
         energy measurements in the calorimeter using the intensive 
         calibration beam program.
In the theoretical predictions, NuTeV is planning to incorporate more
         complete NLO calculations and will investigate the dependence of
         the predictions on factorization scheme
         and various parton distribution function parameterizations. 

While statistical uncertainty dominated in the previous measurement,
         improving the systematic uncertainty would reduce the total 
         error significantly.
We expect the statistical uncertainty  in this analysis would reduce 
         by about a factor of 2.2 combining CCFR and NuTeV data.
The NuTeV is currently planning to present a LO analysis by this summer.

\section{Conclusions}
CCFR measured GLS integral in six $Q^{2}$ bins and extracted the value of
       NNLO $\alpha_{s}$ at the mass of the Z boson.
The measured value of $\alpha_{s}$ is:
\begin{eqnarray}
\alpha_{s}^{NNLO}(M_{Z}^{2})=1.114^{+0.009}_{-0.012}
\end{eqnarray}
which is consistent with the value measured from the structure 
       function analysis~\cite{ex:ccfr-bill} and with world average.
CCFR measured the large-$x$ cross sections in the region $x>0.75$.
Comparisons with various nuclear model seems to point that the FNC model
       along with Fermi-gas and quasi-deutron model describe the
       data well with the exponential fall off parameter $s=8.3\pm0.9$.
The value of $F_{2}$ at $x=1$ is measured to be 
$F_{2}^{\nu Fe}(x=1)=2.1\times 10^{-3}$.

NuTeV has finished data taking in September 1997, running with separate
        beams of $\num$ and $\numb$, and a precision calibration (0.3\%) is
        underway.
Strange sea measurement at the NuTeV experiment is making progress, taking
        advantage of the separate beam and the intensive calibration.
NuTeV expects a LO result in the strange sea measurement by the summer.

\end{document}